\documentclass{PoS}
\usepackage{natbib}

\title{High-resolution polarization imaging of the {\it Fermi} blazar 3C~279}

\ShortTitle{Gamma-rays in 3C~279}

\author{\speaker{B. Rani}$^{1}$\thanks{NPP Fellow} \\
        on behalf of the {\it Fermi}-LAT collaboration \\
        NASA Goddard Space Flight Center, Greenbelt, MD, 20771, USA\\
        E-mail: \email{bindu.rani@nasa.gov}}

\author{S. G. Jorstad \\
        Institute for Astrophysical Research, Boston University, 725 Commonwealth Avenue, Boston, MA 02215, USA \\
        Astron.\ Inst., St.-Petersburg State Univ., Russia}
\author{A. P. Marscher \\
       Institute for Astrophysical Research, Boston University, 725 Commonwealth Avenue, Boston, MA 02215, USA}

\abstract{Ever since the discovery by the {\it Fermi} mission that active galactic nuclei (AGN) produce copious amounts of high-energy emission, its origin has remained elusive. Using high-frequency radio interferometry (VLBI) polarization imaging, we could probe the magnetic field topology of the compact high-energy emission regions in blazars. A case study for the blazar 3C~279 reveals the presence of multiple $\gamma$-ray emission regions. Pass~8 {\it Fermi}-Large Area Telescope (LAT) data 
are used to investigate the flux variations in the GeV regime; six $\gamma$-ray flares were observed in the source during November 2013 to August 2014. We use the 43 GHz VLBI data to 
study the morphological changes in the jet. Ejection of a new component (NC2) during the 
first three $\gamma$-ray flares suggests the VLBI core as the possible site of the 
high-energy emission. A delay between the last three flares and the ejection of a new component (NC3) indicates that 
high-energy emission in this case is located upstream of the 43~GHz core (closer to the black hole). 

}

\FullConference{7th Fermi Symposium 2017\\
		15-20 October 2017\\
		Garmisch-Partenkirchen, Germany}

\begin{document}

\section{Introduction}
{\it Fermi} has brought about a revolution via the discovery of GeV 
emission from active galaxies (AGN). Studies over the past few years suggest 
that rapid $\gamma$-ray flares are produced in regions close to the central 
black hole \citep{rani2013a, rani2013_3c273, rani2014, vovk2016} where the particle acceleration 
is most efficient (in some cases $\gamma$-rays appear to also come from 
larger distances \citep{barnacka2015, barnacka2016, jorstad2010}). 
High-resolution VLBI imaging allows us to probe jet morphology changes  
on scales $\leq$1000~$R_g$, distances where the jets are likely still 
accelerating and are potential sites of $\gamma$-ray emission. 
We present  an investigation of parsec-scale jet morphology 
evolution of the blazar 3C~279 during an episode of extreme $\gamma$-ray flaring 
activity in 2013 -- 2014, using 43~GHz very long baseline interferometry 
(VLBI) images. A detailed study of the event is presented in \cite{rani2017}. 
The $\gamma$-ray bright flat spectrum radio quasar (FSRQ) 
3C~279 \citep[z = 0.538,][]{burbidge1965} has an extremely bright and polarized 
jet pointed close to our line-of-sight at $\leq$2$^{\circ}$ \citep{jorstad2004}.


\section{Results}

\subsection{Gamma-ray variability} 

Gamma-ray photon flux and photon index variations in the source were 
investigated using the {\it Fermi}-LAT \citep[Large Area Telescope,][]{atwood2009} 
data. To explore the GeV variability properties, we generated the constant 
uncertainty (15$\%$) light curve above the de-correlation energy ($E_0$) 
using the adaptive binning analysis method \citep{lott2012}. The light curve is 
produced by modeling the spectra for each time bin by a simple power law, 
N(E) = N$_0$ E$^{-\Gamma}$, N$_0$ : prefactor, 
and $\Gamma$ : power-law index. The details of 
observations and data reduction can be found in \citep{rani2017_3c279}. 

  \begin{figure}[h]
   \centering
\includegraphics[scale=0.65,angle=0, trim=0 0 0 0.0, clip]{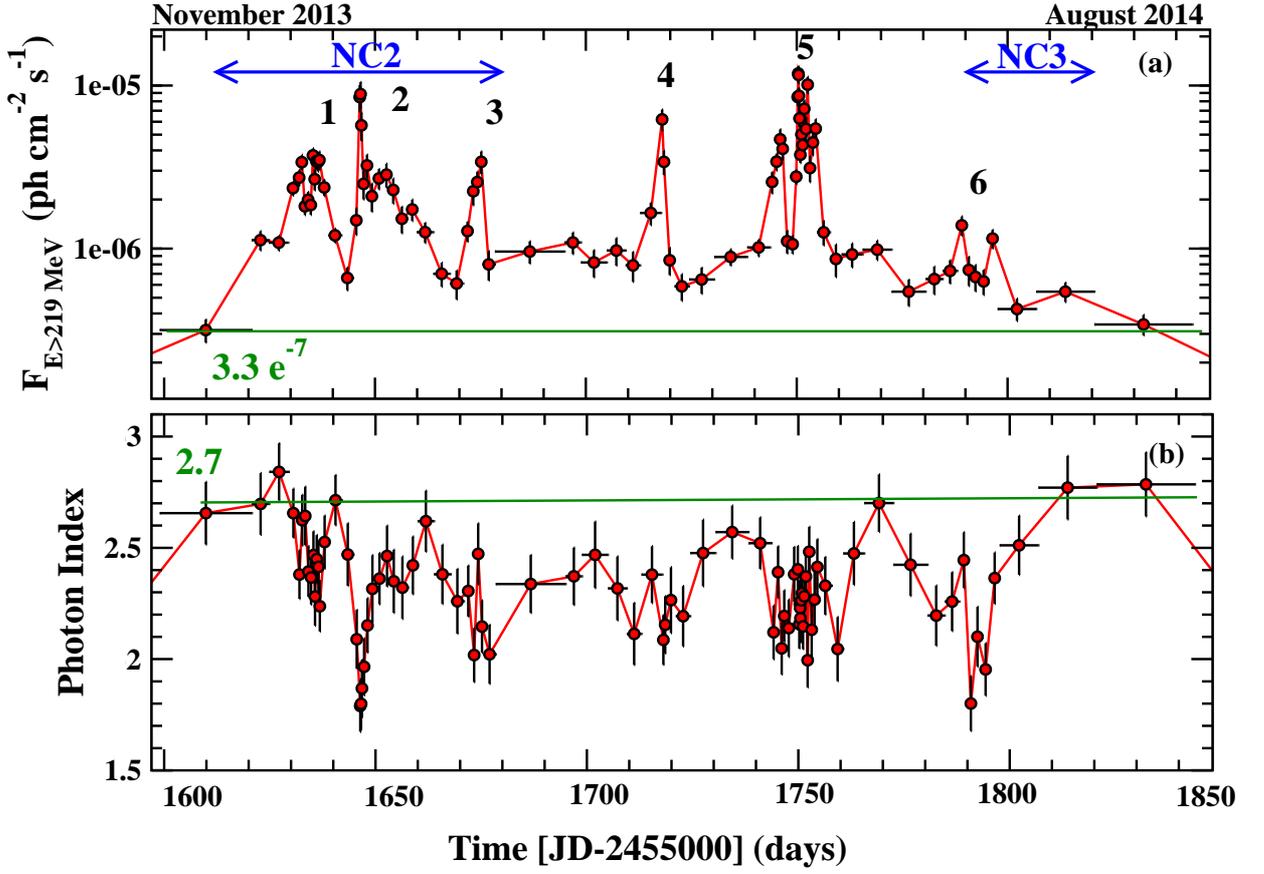}
   \caption{Gamma-ray photon flux and photon index variations observed in 3C~279 
using the {\it Fermi}-LAT at 219~MeV to 300~GeV.  
The rapid $\gamma$-ray flares are labeled as ``1" to ``6". The blue horizontal arrows 
mark the period of ejection of components NC2 and NC3 (see Section \ref{comp_motion} 
for details). }
\label{fig1}
\end{figure}

The source displays multiple modes of flaring activity as can be seen in 
Fig.\ \ref{fig1}. Bright flares, labelled as ``1" to ``6" are superimposed 
on the long-term outburst. An increase in the source brightness from a flux level 
F (E $>$219~MeV) $\sim$ 3.3$\times$10$^{-7}$ photons cm$^{-2}$ s$^{-1}$ 
was observed on  JD$^{\prime}$\footnote{JD$^{\prime}$ = JD - 2455000} $\sim$1610. 
Later in the end of the outburst on  JD$^{\prime}$ $\sim$1850, the source faded to 
the same brightness level. The brightness level of the source before and after the 
outburst is marked via a solid green line in Fig.\ \ref{fig1} (top). Similar variations 
were observed in the $\gamma$-ray photon index as well. The spectrum was softer 
($\Gamma$ = 2.7) before and after the outburst, while significant hardening was observed 
during the flares.

\subsection{Jet morphology variations}
\label{comp_motion}
The 7~mm VLBA observations of 3C~279 were used to investigate the morphological 
changes on parsec-scales. The VLBI data were observed in the course of
the monthly monitoring of bright $\gamma$-ray blazars at 43~GHz 
program\footnote{VLBA-BU-BLAZARS, http://www.bu.edu/blazars}. The standard data reduction 
tasks were performed using the Astronomical Image Processing System (AIPS) and
Difmap (Shepherd 1997). The details of the 
observations and data reduction can be found in \cite{jorstad2017}.

  \begin{figure}
   \centering
\includegraphics[scale=0.6,angle=0, trim=0 0 0 45.0, clip]{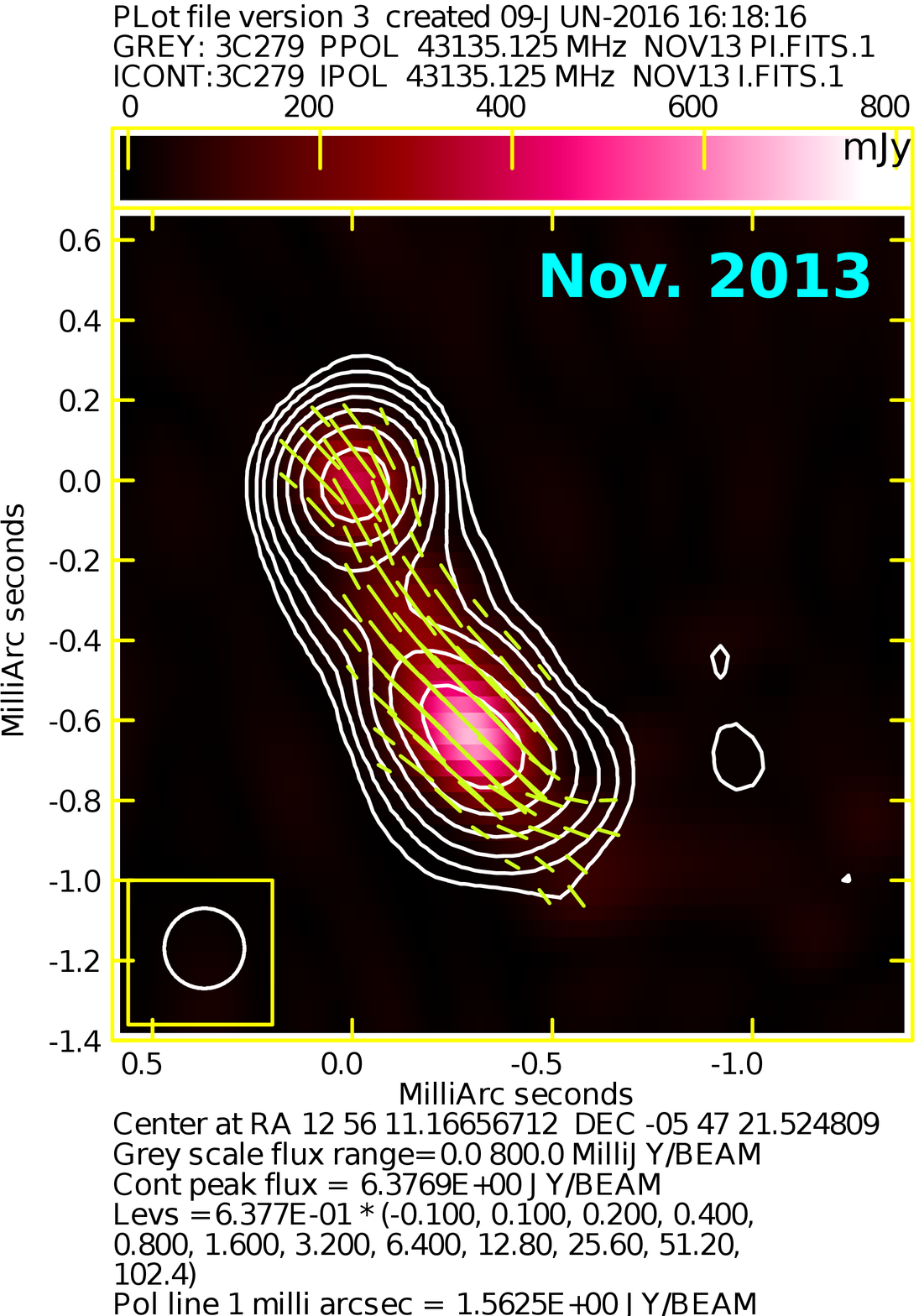}
   \caption{An example composite image of 3C~279 convolved with a beam size of 
0.1~mas (circle in the bottom left corner). The  contours represent the total 
intensity while the color scale is for  polarized intensity image of 3C~279 . The line segments 
(length of the segments is proportional to fractional polarization)  marks the EVPA direction. 
 }
\label{fig2}
\end{figure}

\begin{figure}
   \centering
\includegraphics[scale=0.6,angle=0, trim=0 0 0 0.0, clip]{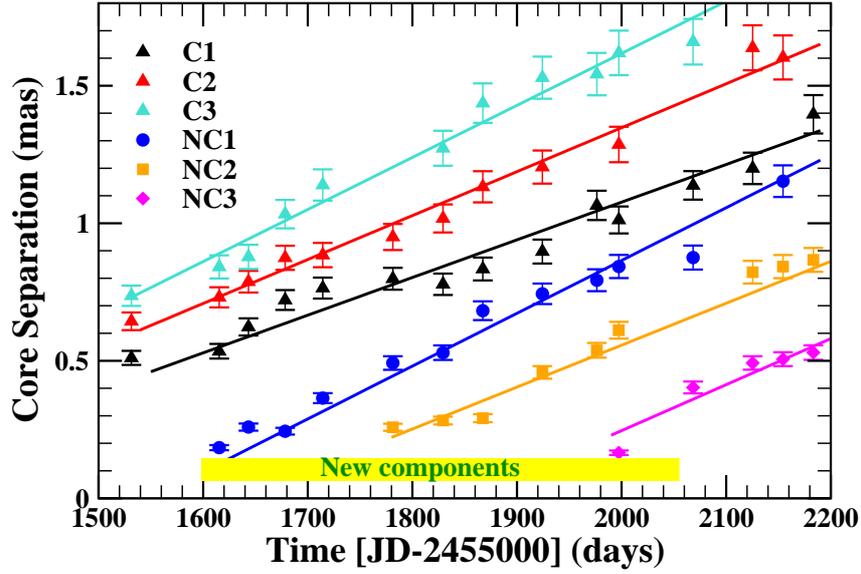}
   \caption{The 43~GHz jet kinematics i.e., temporal evolution of circular Gaussian 
components in 3C~279. The trajectories 
of the components can be well fitted using a linear function (shown using the solid lines). 
 }
\label{fig3}
\end{figure}

Figure \ref{fig2} shows a super-resolved image of the 3C~279 jet convolved 
with a beam size of 0.1~mas. In the source frame, 1~mas corresponds to 6.3~parsec. 
To study the evolution of the jet morphology, we fitted the total intensity of the jet 
using circular Gaussian components. The evolution of the components as a function of 
time is shown in Fig.\ \ref{fig3}. We fix the coordinates of the core to (0,0) 
to study the kinematics of individual components. The solid lines in Fig.\ \ref{fig3} 
shows the best-fitted linear functions, which provide the angular velocity of the 
components. Given the angular speed, one can get an estimate of the apparent speed. 
We found that the new components (NC1 to NC3) have an average apparent speed of 
$\sim$20~c.  Back-extrapolation of the 
components' motion was used to determine  the ejection time or core separation 
time of the new components (NC1, NC2, and NC3). Our calculations suggest that 
NC1 was ejected on JD$^{\prime}$\footnote{JD$^{\prime}$ = JD - 2455000} 1558$_{+15}^{-28}$, NC2 on 1638$_{+40}^{-27}$, and NC3 on 1810$_{+26}^{-19}$~days. 
The ejection periods for NC2 and NC3 are shown in Fig.\ \ref{fig1}.

\subsection{Multiple energy dissipation sites in 3C~279}
Because of the  superposition of multiple modes of flaring activity in 3C~279, 
it quite challenging to establish a connection between component ejection and  
flaring behavior of the source. 
Interestingly,  the component ``NC2" is 
ejected from the core during  the first three (``1" to ``3") $\gamma$-ray flares. 
The horizontal blue arrows in Fig.\ \ref{fig1} mark the ejection period of NC2 and NC3. 
A continuous decay in the source brightness after the next three flares (``4" to ``6") 
is followed by the ejection of NC3. 
Ejection of NC2 during the first three flares  indicates that 
the $\gamma$-ray flares have to be produced either at the VLBI core or very close to the 
core region; otherwise, there should be a delay between the high-energy flares and the 
component ejection, which is not observed. A delay between the last three flares (4 to 6) 
and the ejection of NC3 suggests that $\gamma$-ray are produced upstream of the VLBI core 
(closer to the central engine). Our analysis therefore indicate multiple sites of 
high-energy dissipation in 3C~279.

\section{Future Perspectives} 
VLBI is the  only current technique capable of viewing directly the 
parsec- and subparsec-scale regions of jets in AGNs. Using VLBI, 
we are currently able to achieve an angular resolution of $\sim$20~$\mu$$as$ 
which  makes it the most  promising technique to probe the high-energy 
dissipation sites. The {\it Fermi} mission will continue observing the GeV sky 
at least for next couple of years. The TeV missions are on their way to probe 
the most energetic part of the electromagnetic spectrum.  High-energy polarization 
observations ({\it AMEGO}, {\it IXPE}, etc.) will be of extreme importance in understanding the high-energy dissipation mechanisms. 

\vspace{-0.2in}

\acknowledgments 
\vspace{-0.1in}
{\small
The $Fermi$-LAT Collaboration acknowledges support from a number of agencies and institutes for both 
development and the operation of the LAT as well as scientific data analysis. These include NASA and 
DOE in the United States, CEA/Irfu and IN2P3/CNRS in France, ASI and INFN in Italy, MEXT, KEK, and JAXA 
in Japan, and the K.~A.~Wallenberg Foundation, the Swedish Research Council and the National Space Board
in Sweden. Additional support from INAF in Italy and CNES in France for science analysis during the
operations phase is also gratefully acknowledged.
This research was supported by an appointment to the NASA Postdoctoral Program
at the Goddard Space Flight Center, administered by Universities Space Research Association 
through a contract with NASA.
This study makes use of 43 GHz VLBA data from the
VLBA-BU Blazar Monitoring Program (VLBA-BU-BLAZAR; http://www.bu.edu/blazars/VLBAproject.html), funded
by NASA through the {\it Fermi} Guest Investigator Program. 
The VLBA is an instrument of the Long Baseline Observatory. The Long Baseline Observatory 
is a facility of the National Science Foundation operated by Associated Universities, Inc. 
The BU group acknowledges support from the {\it Fermi} guest investigator grants NNX14AQ58G and 80NSSC17K0649.}


\providecommand{\href}[2]{#2}\begingroup\raggedright
\endgroup

\end{document}